\RequirePackage{fix-cm}
\documentclass[twocolumn]{svjour3}          
\smartqed  
\usepackage{graphicx}
\usepackage[inline, shortlabels]{enumitem}
\PassOptionsToPackage{hyphens}{url}\usepackage{hyperref}
\usepackage{microtype}

\begin{document}

\title{Large Process Models
}
\subtitle{A Vision for Business Process Management in the Age of Generative AI}


\author{Timotheus Kampik \and Christian Warmuth \and Adrian Rebmann \and Ron Agam \and Lukas N.P. Egger \and Andreas Gerber \and Johannes Hoffart \and Jonas Kolk\and Philipp Herzig \and Gero Decker\and Han van der Aa\and Artem Polyvyanyy \and Stefanie Rinderle-Ma \and Ingo Weber \and Matthias Weidlich}


\institute{Timotheus Kampik \and Christian Warmuth \and Adrian Rebmann \and Ron Agam \and Lukas N.P. Egger \and Andreas Gerber \and Johannes Hoffart \and Jonas Kolk\and Philipp Herzig\and Gero Decker \at
              SAP, Germany \\
              \email{first.last@sap.com}           
           \and
           Han van der Aa \at
              University of Vienna, Austria \\
              \email{han.van.der.aa@univie.ac.at}
              \and
              Artem Polyvyanyy \at
              University of Melbourne, Australia \\
              \email{artem.polyvyanyy@unimelb.edu.au}
              \and
              Stefanie Rinderle-Ma \at
              Technical University of Munich, School of CIT, Germany \\
              \email{stefanie.rinderle-ma@tum.de}
              \and Ingo Weber \at
              Technical University of Munich, School of CIT \& Fraunhofer Gesellschaft, Germany \\
              \email{ingo.weber@tum.de}
              \and Matthias Weidlich \at
              Humboldt-Universität zu Berlin, Germany \\
              \email{matthias.weidlich@hu-berlin.de}
}

\date{Received: date / Accepted: date}

\maketitle
\sloppy
\begin{abstract}
The continued success of Large Language Models (LLMs) and other generative artificial intelligence approaches highlights the advantages that large information corpora can have over rigidly defined symbolic models, but also serves as a proof-point of the challenges that purely statistics-based approaches have in terms of safety and trustworthiness. As a framework for contextualizing the potential, as well as the limitations of LLMs and other foundation model-based technologies, we propose the concept of a Large Process Model (LPM) that combines the correlation power of LLMs with the analytical precision and reliability of knowledge-based systems and automated reasoning approaches.
LPMs are envisioned to directly utilize the wealth of process management experience that experts have accumulated, as well as process performance data of organizations with diverse characteristics, e.g.,\ regarding size, region, or industry. In this vision, the proposed LPM would allow organizations to receive context-specific (tailored) process and other business models, analytical deep-dives, and improvement recommendations. As such, they would allow to substantially decrease the time and effort required for business transformation, while also allowing for deeper, more impactful, and more actionable insights than previously possible. We argue that implementing an LPM is feasible, but also highlight limitations and research challenges that need to be solved to implement particular aspects of the LPM vision.
\end{abstract}

\section{Introduction}
\label{sec:intro}
The recent success of transformer architectures has positioned Large Language Models (LLMs) at the frontier of artificial intelligence research and applications. The general idea of LLMs and other so-called \emph{foundation models} is to use a large body of not explicitly labeled data for training, to then infer ``statistically plausible'' outputs (in a given modality, such as text or image), given an input. Promising applications of LLMs are emerging in the enterprise software industry; while general-purpose LLMs can already augment day-to-day knowledge work such as copywriting, specialized models are trained for domains such as software engineering~\cite{10.1145/3524842.3528470} and finance~\cite{wu2023bloomberggpt}. However, as LLMs are statistics-based tools that re-use large corpora of often poorly curated, human-generated text, their behavior is unpredictable, at times not desirable, and frequently illogical. This limits the applicability of (plain) LLMs in many business contexts. In particular in Business Process Management (BPM) and process intelligence, where decisions have critical implications for business operations, the raw and astonishing correlation power of deep learning is insufficient as a standalone facilitator of reliable, trustworthy, and actionable intelligence. To facilitate intelligence with the aforementioned properties, an integration of LLMs (or more broadly, foundation model-based approaches) with symbolic data management (such as knowledge graphs) and automated reasoning methods is required.

In this paper, we propose Large Process Models (LPMs)\footnote{Large process models are not to be confused with \emph{local process models} (also typically abbreviated using the \emph{LPM} acronym), which aim to describe behavior that frequently occurs in an event log that a process has generated in \emph{local}, somewhat \emph{small} patterns~\cite{TAX2016183}.} as a central conceptual framework for software-supported BPM in the era of generative AI, with the overall objective to provide a balanced, feasibility-oriented discussion of the expected impact of foundation models on BPM software.
We ground LPMs in the state-of-the-art of the two research areas (Section~\ref{sec:background}) to then provide a motivation for the LPM concept from different perspectives (Section~\ref{sec:motivation}).
Drawing from existing research, we then assemble the LPM from both nascent and well-established components (Section~\ref{sec:lpm}).
We discuss the application potential of LPMs (Section~\ref{sec:potential})
and argue for their technical feasibility \emph{to a certain extent}, while
also highlighting substantial risks and challenges from both academic and
practical perspectives (Section~\ref{sec:feasibility}).
Finally, we discuss concepts related to LPMs, as well as LPMs beyond LLMs (Section~\ref{sec:discussion}), before we conclude the paper (Section~\ref{sec:conclusion}).

\section{Background}
\label{sec:background}
BPM is a professional discipline and research area that is concerned with ensuring that organizations run as desired and achieve their competitive and societal objectives.
In the context of BPM, software plays an important role, as organizations rely on it not only for the execution of processes, but also for process design and analysis.
BPM as a research discipline, while multi-disciplinary, is often seen through the lens of applied computer science, e.g.,\ in the context of foundational approaches to process modeling~\cite{DBLP:books/sp/Weske19} and data-driven process analysis (process mining)~\cite{DBLP:books/sp/Aalst16}.
Here, ``formal'' modeling languages, such as Petri nets, play a crucial role to facilitate reasoning and decision-making about processes.
Traditionally, BPM as a field and, in particular, BPM software, rely on symbolic approaches to computer science, many of which are logic-based and may consequently be considered ``good old fashioned'' symbolic AI.
For instance, the core of process mining is based on symbolic data management and temporal reasoning approaches.

With the advent of deep learning, BPM research has adjusted its course and increased the uptake of a variety of Machine Learning (ML) approaches.
As a reaction to this trend, the BPM community has provided a vision of \emph{AI-augmented BPM}~\cite{2023-Dumas}.
According to this envisioned approach, subsymbolic AI methods are not used to replace human or symbolic reasoning in crucial tasks, but rather to support human and machine decisions and actions, e.g.,\ in order to facilitate human control with less effort, while still allowing for strong, symbolic guarantees.
The ultimate goal of AI-augmented BPM is making business processes ``adaptable, proactive, explainable, and context-sensitive''~\cite{2023-Dumas}.
Its two key elements, human-control and the integration of symbolic (logic/reasoning-based) and subsymbolic (statistics/learning-based) AI approaches are well-established research directions in the AI community: the fusion of symbolic and subsymbolic AI is well-known since the turn of the century as \emph{neuro-symbolic AI}~\cite{garcez2002neural} and is currently re-surging, for example in the context of knowledge graphs and the Semantic Web~\cite{10.1145/3586163}.

Recently, the rise of so-called generative AI, enabled by the transformer neural network architecture~\cite{DBLP:conf/nips/VaswaniSPUJGKP17}, has fueled new expectations regarding the application potential of artificial intelligence, not least in business and BPM contexts.
Most prominently, software products such as ChatGPT allow users to engage in dialogues with LLM-based systems that then produce statistically plausible content given a user's request, based on the large corpora of content that the systems have been trained on.
There is substantial interest regarding generative AI in BPM research, as well as in industry.
For example, recent research provides first insights into the potential that generative AI has for process mining (in particular: query generation and direct question answering based on event logs~\cite{berti2023abstractions}).
Also, emerging research lines explore creating models of processes (process model generation~\cite{fill2023conceptual} and task list extraction from text~\cite{klievtsova2023conversational}), as well as related conceptual models such as Unified Modeling Language (UML) models~\cite{chatGPTUML,fill2023conceptual}, and academic proposals for prompt engineering for BPM and enterprise modeling have been introduced~\cite{10.1007/978-3-031-34241-7_1,10.1007/978-3-031-48583-1_1}.
What is lacking so far is a holistic overview of how generative AI can facilitate BPM more broadly, and how a systematic perspective on the interplay with existing technologies can be developed.

\section{Motivation}
\label{sec:motivation}
The concept of a \emph{large process model} can be motivated from a dual perspective.

Intuitively, the increasing interest in LLMs in a broad range of domains calls for their holistic positioning in the context of BPM. Considering the prevalence of process \emph{models} as tools for process analysis as well as execution artifacts, the term \emph{large process model} can refer to the application of LLMs in order to produce models of processes in the broader sense.

However, beyond this simplistic analogy, we view ``large process model'' more literally as an alternative to the small, hard-wired, and specific process models that are used today (think of a BPMN\footnote{Business Process Model and Notation, an open standard for modeling business processes~\cite{omg2011bpmn}} model and an associated DMN\footnote{Decision Model and Notation, an open standard for modeling business rules~\cite{omgdmn}} rule base) that goes beyond the mere application of LLMs.

In Natural Language Processing (NLP), the term LLM refers to statistical models of natural language that, based on the large corpora of text data they have been trained on, predict next plausible \emph{tokens} (basic units of text) given an input string~\cite{zhao2023survey}. Initially, LLMs applied relatively simple statistics-based approaches, which have been replaced by neural networks during the 2010s; today's LLMs are typically \emph{Generative Pre-trained Transformers} (GPT), utilizing the corresponding transformer neural network architecture that was specifically designed for NLP tasks (although it is more generally applicable).
Predecessors of statistics-based NLP approaches such as the early LLMs were formal/logic-based models of language. These symbolic models attempt to precisely define the meta-model of human language(s) and allow for the instantiation of these meta-models in particular contexts. However, considering the complex, nuanced, and dynamic nature of human language, logic-based language models are insufficient for handling most NLP tasks and are now assumed to be primarily applicable in conjunction with ML-based tools such as LLMs~\cite{nlp-neurosymbolic}. Analogously, BPM used to have a strong symbolic, model-driven focus, in particular in academia, which is reflected by classical textbooks on the topic~\cite{DBLP:books/sp/Weske19,DBLP:books/sp/DumasRMR18}. The assumption was that \emph{imperative} process models allow business experts to specify how organizations run in a precisely defined and automatable manner, following a model-driven development approach.
In contrast, more permissive, \emph{declarative} approaches to modeling business processes have not yet achieved mainstream maturity, possibly due to challenges regarding unified representation and user-friendliness of creating and managing models; this is evidenced by the relative lack of adoption of the CMMN standard\footnote{Case Management Model and Notation, an open standard for the declarative specification of business processes~\cite{omg2016cmmn}}, which at least some specialized vendors with deep expertise in the field regard as a failure\footnote{Cf. \url{https://camunda.com/blog/2020/08/how-cmmn-never-lived-up-to-its-potential/}, \emph{accessed at 28-03-2024}}.

Even for imperative process models, and although a mature technology ecosystem of business process execution engines exists, the direct deployment of models for execution has remained a niche approach to process automation. An industry assumption is that the level of business experts' technology literacy and maintenance effort required for model-driven development is so high that only very large and mature organizations can benefit from it, and typically only in the most business-critical parts of their operations; for others, standard software or traditional custom development remain more viable. Thus, similarly to the natural language case, the dream of a perfect symbolic model remains an ambition that is rarely achievable.
The following  two examples highlight this issue and exemplify two vastly different BPM scenarios (highly customized and largely standardized processes). \\

\noindent \textbf{Compliance checking process capabilities for the finance domain.} Large financial institutions typically want to be in full control of their business process specification and execution and hence apply model-driven development stacks with open source or self-built business process and rule execution engines, utilizing modeling notations such as BPMN and DMN\footnote{Notably, a well-known Wallstreet bank maintains their own DMN engine: \url{https://github.com/goldmansachs/jdmn} (\emph{accessed at 28-03-2023}).}. A key use case within the domain is ensuring regulatory compliance while maximizing business agility. However, the model and rule bases needed for executing the corresponding checks and integrating them into core business operations are very large and the maintenance effort is immense. Symbolic models and rules are, even if correct from a ``logical'' perspective (\emph{object-level}), prone to be dated, inconsistent, or incorrectly modeled from a domain perspective (\emph{meta-level}). Hence, substantial human effort, as well as very particular expertise at the intersection of technology and the specific business domain, is required for maintenance and continuous improvement. \\

\noindent \textbf{Generic purchase-to-pay process capabilities.} Enterprise software vendors scale generic purchase-to-pay (procurement) process capabilities across thousands of organizations using standard software. Changing and customizing the software is often effortful and introduces risks of unintended side effects that need to be mitigated. Hence, the degree of customization needs to be a carefully deliberated trade-off. Making the right decision about the scale and direction of customization requires digging into data and knowledge silos; the available data typically tells only a part of the story, and the most useful knowledge is typically distributed across different sources, hard to find, and not available in a machine interpretable format.

At the same time, the utilization of traditional statistics- and ML-based approaches (non-GPT methods) in a BPM context poses substantial challenges, in particular due to the following key issues:
\begin{itemize}
    \item BPM is knowledge-intense and classical statistical inference approaches struggle with the utilization of organizational knowledge, in particular considering that this knowledge is typically not available in a structured, well-maintained and easy-to-process form.
    \item Deep learning approaches that require training neural networks from scratch are extremely costly to scale; training for a particular organizational context is often not feasible given that business processes typically drift with time, and continuous re-training is required.
    \item Reinforcement learning approaches that can potentially further systematize and partly automate the continuous improvement of business processes~\cite{DBLP:journals/is/SatyalWPCM19} depend on knowledge for checks and balances. Learning by action comes at a cost, in particular in scenarios where the distribution of utility generated by rewards is time sensitive: in a BPM context, bad rewards tend to come in late, e.g., in the long tail of process instances that eventually turn out to not terminate as intended.
\end{itemize}
Consequently, the collection of knowledge and data across processes, organizations, industry verticals, and process variants in an LPM in the broader sense can enable a substantial step forward: instead of relying on one specific, yet simplistic and incomplete model, all models are utilized to the extent they are useful in order to manage a particular process (or variant or instance thereof).

\section{Large Process Models}
\label{sec:lpm}
To advance a holistic viewpoint on the technological foundations of BPM software in the age of generative AI, we propose the concept of a Large Process Model (LPM).

An LPM is envisioned as a neuro-symbolic software system that integrates process management knowledge accumulated by experts and precise data on how organizations run their processes with generative AI approaches and statistical as well as symbolic inference methods, thus fusing process data and knowledge. Given process data in an event log or relational format\footnote{Additional data, such as survey data reporting on customer or employee satisfaction can be utilized; still, we assume that traditional tabular data representing process execution traces forms the core input for our data-driven inferences.}, the LPM automatically identifies the domain of a specific process as well as the context of the organization that runs it, to then generate
insights and action recommendations, using a collection of tools for process design, analysis, execution, and prediction.
As organizational context, process data alone is sufficient, but additional information, e.g., in the form of process models or unstructured documents, can be automatically ingested in order to augment context-specific LPM capabilities. LPM knowledge is partially encoded in an LLM and partially managed as symbolic \emph{process atoms}, which are models and query templates generated by an ensemble of deep learning techniques and special-purpose algorithms. Depending on the BPM task at hand, the LPM is instantiated from the general framework presented. The implementation of the tasks is not necessarily hard-coded, but can be tackled more flexibly utilizing agent-based approaches, i.e., reasoning loops that have been at the center of AI research for decades and are now to some extent applied in the context of LLMs~\cite{yao2023react,wang2023planandsolve}\footnote{Let us note there that the way the notion of an ``agent'' is used in the context of LLMs is subject to community debate and that many of the nascent agent system proposals and prototypes do not yet make use of the comprehensive planning, reasoning, and (reinforcement) learning capabilities that have been devised over the last decades.}.

\begin{figure*}[ht]
\centering
\includegraphics[width=0.75\textwidth,trim=22cm 0cm 23cm 0cm,clip]{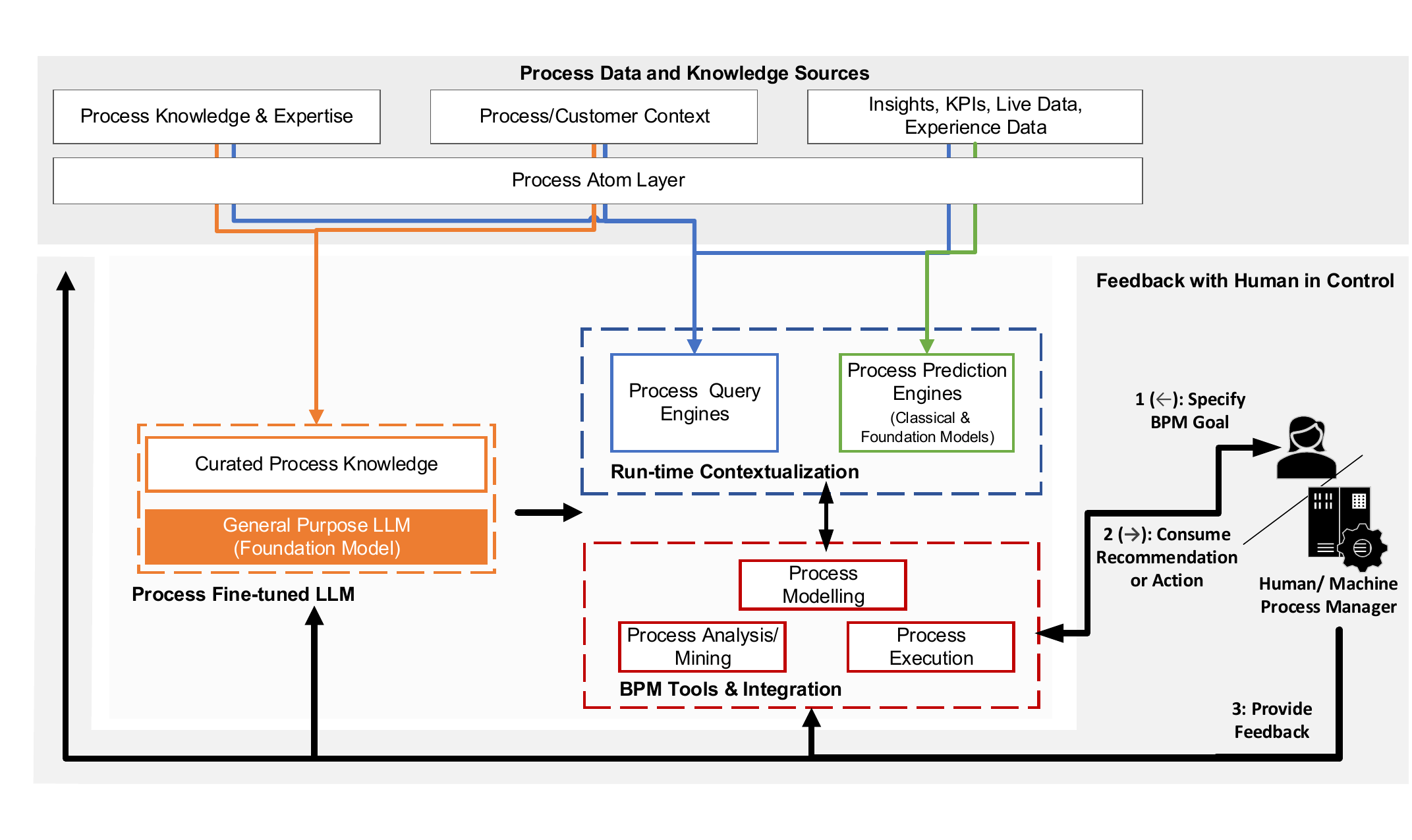}
\caption{A conceptual architecture of an LPM.}
\label{fig:lpm}
\end{figure*}

The LPM consists of the following key components (Figure~\ref{fig:lpm}). \\

\noindent \textbf{Process data and knowledge sources.}
    Process data and knowledge is provided to train ML models, feed symbolic algorithms, and to serve humans directly.
    On the technology side, structured knowledge is persisted in knowledge graphs (or is managed using other knowledge-based technologies) and, in the case of language embeddings, in vector stores; unstructured data is typically available in a multitude of formats, whereas tabular data is available in traditional relational databases. Note that here, we do not aim to utilize knowledge graphs for the representation of event data, unlike recent process mining approaches~\cite{Fahland2022}, mostly due to scalability concerns; instead, the term \emph{knowledge graph} can be understood as a pars pro toto for semantic technologies persisting process \emph{model} knowledge.
    Conceptually, we can divide data and knowledge sources into generic knowledge about a process or its application domain (industry vertical), customer and context-specific knowledge, and execution data.
    Here, knowledge is often distilled from data on a continuous basis, e.g.,\ in the case of benchmarks.
    In order to utilize the content of heterogeneous knowledge sources, the LPM proposal features the notion of a \emph{process atom layer}: process atoms are facts about a process (or relationships between facts) that are atomic in the sense that they cannot be split into smaller parts without losing their business meaning.
    For example, when checking the violation of the rule ``only if the order amount exceeds \$1,000 then an approval occurs'', the logical implication can be split into ``smaller'' propositions, but it makes little sense to do so from a business user perspective.
    Breaking down process knowledge into process atoms and mining process atoms from unstructured information or data closes the gap between the natural-language focus of LLMs and the need for representations that are executable, in particular as queries on tabular process data and symbolic process knowledge.
    Intuitively, process atoms can be considered equivalent to declarative process models or \emph{constraints}, whose (interactive) discovery from event logs~\cite{5949297} and integration with imperative modeling languages such as BPMN~\cite{10.1007/978-3-319-19069-3_6} has already been studied by the scientific community and that can, for example, be sourced using LLMs for the purpose of detecting so-called semantic anomalies~\cite{DBLP:journals/is/AaRL21,caspary2023does}, i.e., of behaviors that are unusual given the knowledge at hand and hence possibly undesirable.
 \\

\noindent \textbf{Process fine-tuned LLM.} Both structured knowledge and unstructured information are used to fine-tune an LLM. Notably, an LPM is not an LLM trained on domain-specific knowledge. For efficiency and flexibility, the assumption is that fine-tuning and contextualization in prompts is preferable over training from scratch. Fine-tuning can take place across multiple dimensions.
    For example, an LLM may be fine-tuned for:
    \begin{itemize}
        \item process management in general: with specific terminology and general knowledge about BPM;
        \item a particular process vertical;
        \item a specific region and its norms and regulations;
        \item a specific organization.
    \end{itemize}

\noindent \textbf{BPM tools and integration.}
    Considering that BPM is, in many aspects, a precise discipline, in which properties such as reliability and trustworthiness play an important role, it would be naïve to assume that an LLM can fully replace existing tooling.
    For example, documentation on how processes run must be interpretable and manageable in a systematic manner, and in most cases, process performance and conformance assessments must yield ``hard'' guarantees and not approximate guesses.
    Hence, an LPM must rely on classical process management tooling to combine the benefits of LLMs with the primarily symbolic data management-based classical BPM tools and algorithms, in particular for process modeling, analyses, and execution, whereas predictive capabilities can be provided by statistics-based models (see below).
 \\

\noindent \textbf{Run-time contextualization.}
    The backbone of the aforementioned tools is formed by process knowledge and process data query engines.
    These engines provide the basic ``plumbing'' for enterprise-grade BPM, such as access control management capabilities, as well as interfaces for human users.
    Accordingly, the infrastructure for process knowledge and process data querying needs to be provided and augmented to work in interplay with LLM-based inferences.
    For example, an LLM could generate plausible hypotheses about a process based on its execution data, and these hypotheses can then be tested in a rigorous manner using traditional symbolic and statistical inference algorithms\footnote{In general, the fusion of LLMs and knowledge-based approaches as described here is often referred to as retrieval-augmented generation~\cite{NEURIPS2020_6b493230}.}.
    For predictive analysis and simulation, a collection of ML models is utilized. Here, best-in-class models for a given task or ensembles can be used. Given the success of foundation models in the domain of natural language processing, foundation models trained on business process execution traces may be a promising augmentation of more traditional ML models.
 \\

\noindent \textbf{Inherent feedback mechanism with human in control.}
    In order to ensure that the inferences drawn by the LPM are indeed useful, feedback mechanisms with the human in control are proposed. Feedback can either be provided fully by machines, e.g., by automatically determining, based on heuristics, whether a generated and executed query yields relevant results.
    However, in many cases, the final arbiter can be expected to be a human, who needs to be involved when designing process models, interpreting the business meaning of key performance indicators and other quantitative insights, and approving the deployment of process changes to (enterprise) information systems. Accordingly, the feedback loop interpretation needs to support adequate technical feedback capabilities that enable reinforcement learning, but also consider human-computer-interaction factors to facilitate rational decision-making. The interactive feedback and inference loops can be orchestrated by light-weight workflow engines or reasoning loop architectures; due to their stochasticity, we assume that LLMs do \emph{not} play a role in orchestration.

The conceptual LPM architecture features the following iterative user-flow (also see Figure~\ref{fig:lpm}), for either a machine or human user aiming to accomplish a BPM task\footnote{Let us note that the focus of the LPM is on business process \emph{management}, i.e.,\ on making sure an organization runs in a desirable manner, and not on the execution of particular process instances on a case-by-case basis.}.
\begin{enumerate}
    \item \textbf{Specify (BPM) goal.} The user specifies the BPM goal they desire to achieve alongside boundary conditions. Such a goal can be relatively straightforward, as in ``give me the most important configuration changes that help reduce cycle time in my order-to-cash process,'' or it could be very ambitious, as in ``change the order-to-cash process implementation so that the cycle time is reduced without negatively affecting other Key Performance Indicators (KPIs).'' In case the goal is specified by a machine user, we expect that it is triggered by a context in which a human user is situated, e.g., from a control panel accessed by a human user, in which an order-to-cash process with sub-par cycle time is summarized.
    \item \textbf{Consume recommendation or action.}
    The specified goal is turned into a prompt, based on which the process fine-tuned LLM produces the desired content and triggers queries to other subsystems. Upon query response, the initial prompt may be further refined, or additional queries may be generated. For example, the LPM may search for process variants for which the cycle time is particularly high to then identify potential root causes based on correlation analysis and textual knowledge; for this, repeated querying of the event log and evaluation of the returned query results are required.
    \item \textbf{Provide feedback.} Based on the results returned by the LPM, feedback is provided, either by a human or a machine (the latter of which may also be an LPM component). For example, if the process data query engine returns an empty set, this can be considered negative feedback in many contexts.
    Human feedback is necessary in more nuanced cases, i.e.,\ to provide context that does not exist within the boundaries of the purely technical system. Human feedback could assess some action recommendations as particularly useful, while marking others as false positives. For instance, in the context of a process change recommender system, a human expert may have a better overview of the social effects, risks, and costs of organizational change and assess some recommendations as not viable, because they are unlikely to affect meaningful process change or because carrying them out is too risky or too costly.
    The feedback can then be used to fine-tune the LLM, to generate labeled data, and to train more classical recommender systems that utilize reinforcement learning-based approaches such as contextual multi-armed bandits~\cite{lu2010contextual} to continuously improve the LPM.
\end{enumerate}

\section{How LPMs Can Facilitate BPM}
\label{sec:potential}
Below, we provide an outline of how LPMs can facilitate BPM. We point to specific research results that provide additional context and preliminary or partial evidence of potential feasibility\
to then provide a more nuanced discussion of feasibility in Section~\ref{sec:feasibility}.
\begin{enumerate}
    \item \textbf{Reduction of effort \& expertise required for knowledge-based BPM tasks.}\label{item:effort}
    Managing business processes is knowledge-intense work, requiring both in-depth expertise with respect to specific tools and skill sets, such as process modeling notations and process data query languages and access to and a good grasp of the knowledge and data that exists about a particular process, typically in a highly complex organizational context. Hence, human BPM experts (individuals or teams) must have a high level of technical and socio-professional skills, as well as substantial experience within a particular organization: the entry bar for successfully running a BPM initiative is high.
    LPMs can lower this entry bar by \emph{i)} making it easier to find and automatically present information in the context in which it is useful and tailored exactly to the user's skill level and expertise; \emph{ii)} turning unstructured and semi-structured information into models and queries, thus requiring less detailed and formal knowledge of languages for process design and analysis; \emph{iii)} enriching and extending contextual process information based on logically inferred or statistically plausible facts.
    In the context of this broader objective, we envision, for example, the following specific LPM-capabilities:
    \begin{itemize}
        \item Turning natural language text into process models and queries (of process models and data);
        \item Enhancing process models and queries based on natural language feedback;
        \item Recommending (changes to) process models and queries based on natural language context;
        \item Scaling generic insights derived from process data across organizations by auto-generating templates from commonly executed queries and instantiating them automatically in a given context.
    \end{itemize}
    Here, we expect that the emerging notions of \emph{conversational process modeling}~\cite{klievtsova2023conversational} and \emph{conversational process mining}, whose human-in-the-loop-level feasibility is to some extent supported by recent research results~\cite{klievtsova2023conversational,berti2023abstractions,DBLP:journals/corr/abs-2307-09923} will become a reality and find their way into production-grade BPM software in an incremental manner over the coming years.
    \item \textbf{Improvement of \emph{process observability}.}\label{item:observability}
    A key challenge in BPM is data handling; the key method exemplifying this is process mining, which is widely considered a cornerstone of modern process analysis. Process mining uses event logs that have been extracted from enterprise systems as input data; these event logs are typically not readily available and generating them as the result of Extract-Transform-Load (ETL) pipelines is known to incur substantial efforts~\cite{DBLP:conf/bpmds/KampikW22}. Even when event logs are generated, they only contain a small subset of the process data that exists in an organization, e.g., because not all relevant IT systems can be accessed or because substantial parts of the process are executed through informal channels (and are hence not recorded in database tables). Also, interpreting what occurs in the event log is typically not trivial, for example, because the business meaning of events is not always clear, which increases the risk of misinterpretations.
    From an industry perspective, we summarize these challenges under the umbrella of \emph{process observability}, which refers -- somewhat analogously to data observability in distributed systems~\cite{10.1007/978-3-030-33702-5_3} -- to the extent to which a process is correctly and completely observed and understood, given the (business) objective at hand\footnote{For an informal introduction to process observability see: \url{https://blogs.sap.com/2022/09/16/what-is-business-process-observability-and-why-does-it-matter/}, \emph{accessed at 28-03-2023}.}. We claim that process observability tends to remain relatively low when relying merely on one analysis method such as modeling or (event log-based) mining. By fusing the knowledge and data from a wide variety of sources, the LPM can potentially increase process observability.
    For example, the following LPM capabilities can potentially facilitate process observability:
    \begin{itemize}
        \item Turning vast amounts of unstructured, informal process knowledge into actionable models, and queries by setting up a pipeline that systematically searches through organizational knowledge silos\footnote{This relates to the previous broad objective, but imagine a larger scale, as well as a more systematic approach.}; here, existing approaches to interactive process discovery that utilize domain knowledge can be employed~\cite{10.1007/978-3-030-00847-5_19,SCHUSTER2022103612}, where the domain knowledge is then (partially) sourced using LLMs.
        \item Utilizing natural language information to discover data sources in large information system landscapes and to recommend ETL scripts/queries for extracting relevant data, advancing and applying existing research on NLP for ETL~\cite{DBLP:journals/widm/DibaBWW20};
        \item Enabling data-driven forecasting and analysis based on foundation models, avoiding the training of specific models for a specific organization's process(es).
    \end{itemize}
    \item \textbf{Convergence of process design, execution, and analysis.}\label{item:lifecycle}
    Finally, the LPM may eventually help organizations to advance towards truly continuous automated process improvement where process design, execution, and analysis converge. The idea of autonomous business process improvement has already been studied in-depth in the context of business process execution engines, which are augmented with reinforcement learning capabilities that over time learn the best process variant for a given context~\cite{DBLP:journals/is/SatyalWPCM19}. This highlights the practicality of the general idea, albeit in an engineering setup resembling immaculate model-driven development that is typically not achievable in the context of real-life BPM deployments and implementations. In reality, process models are typically not deployed with the click of a button; instead, complex and knowledge-intense configuration workflows must be executed to finally trigger an update.
    More broadly, the challenge of turning process analysis outcomes into specific \emph{actions} that change how a process runs is emerging as a key challenge in research~\cite{10.1007/978-3-031-46846-9_15} as well as in practice.
    By making knowledge readily available in a given context, the LPM can make these brittle and human work-intense configuration flows more agile and resilient. Also, if realized, better simulation and prediction capabilities can substantially decrease the risk of deploying process changes.
    Here, we envision the following LPM capabilities, in the spirit of AI-augmented BPM~\cite{2023-Dumas} (for example):
    \begin{itemize}
        \item Matching process analysis insights to potential actions and their assumed consequences;
        \item Fusing data and knowledge to holistically assess the implications and risks of particular process change actions;
        \item Continuously assessing deployed process changes and fine-tuning them for optimal performance.
    \end{itemize}
    Still, we claim that human control and oversight should always play a role in continuous process improvement, to avoid that machines get stuck in local optima or optimize towards clearly undesirable process behavior.
\end{enumerate}
The following example highlights the aforementioned three high-level potential benefits of LPMs.
Consider a purchasing organization that is just getting started with business process management and wants to adopt a data-driven approach right away.
Process-level KPIs, extracted directly from the enterprise system's relational database, have indicated that the process performs poorly in terms of cycle time; for an in-depth analysis, the application of process mining is required.
Based on unstructured system documentation, the LPM suggests which tables to extract the data for the purchase-to-pay process from, recommending a configuration of the ETL connector that merely needs minor adjustments (Benefit~\ref{item:observability}).
After the data is ingested, an automated data analysis is executed.
The organization's ERP system executing the process is highly customized and integrates with self-built sub-systems and services.
Hence, there is no exact reference process model that applies.
Based on a large collection of (potential) reference process models, as well as based on organization-specific textual documentation, the LPM generates a set of queries for conformance checking, as well as for quantitative analyses, executes them, and ranks their results and basic business interpretation by relevance for the extracted event log (Benefit~\ref{item:effort}).
For example, the conformance check may show that \emph{maverick buying} (purchasing without a requisition) occurs frequently, leading to increased time to process completion and compliance risks for purchase order amounts larger than 10,000\$.
The results are then linked to action recommendations, based on ``historic'' process management knowledge (Benefit~\ref{item:effort}), as well as data and models of other organizations' purchase-to-pay processes (Benefit~\ref{item:observability}).
Finally, the most promising action recommendations are applied to the system configuration, where they are (semi-automatically and carefully) first shadow-tested and then piloted, to be finally either discarded or fully applied to the entire production system (Benefit~\ref{item:lifecycle}).
In our maverick buying example, possible changes could be the addition of a pre-approval step for large order amounts, or the ``hard'' enforcement of the ordering of activities for all or some cases that exceed purchase order amounts of 10,000\$.

\section{Feasibility and Challenges}
\label{sec:feasibility}
We envision that LPMs will emerge in an iterative manner, which will help ensure that the capabilities provided live up to ethics, quality, and compliance expectations. Below, we provide a three-step outline of how LPMs can potentially evolve and mature.
We start with capabilities that we consider generally feasible given the state of the art (Step~\ref{step1}) and then move, via capabilities that pose substantial challenges whose solutions are still nascent (Step~\ref{step2}), to a ``blue sky'' vision that focuses more on what is intuitively desirable than on what is feasible (Step~\ref{step3}).
At each step, we argue, based on the scientific literature, for the feasibility of the capabilities or the lack thereof and highlight some of the research challenges that we see\footnote{Obviously, the list of research challenges is non-exhaustive, i.e., it can serve as an opinionated starting point.}.

\begin{enumerate}
    \item \textbf{Augmenting modeling and analysis with contextualized knowledge.}\label{step1}
    The first step towards the LPM vision is the utilization of business process knowledge that would otherwise either not be findable or could not be structured in a way that allows for partially automated analysis with the human in the loop.
    Here, the two main capabilities are LLM-augmented process modeling and mining.
    Even before the emergence of LLMs and foundation models, a substantial line of research has focused on extracting process models from unstructured information and in particular text~\cite{bellan2021process,klievtsova2023conversational,DBLP:conf/caise/SaiWFR23,van2019extracting} and, conversely, on turning symbolic process models into natural language-based artifacts~\cite{DBLP:journals/tse/LeopoldMP14}.
    These models can be either imperative, like classical BPMN models, or declarative, like constraint-based queries that are executed on an event log, e.g., for conformance checking purposes.
    Furthermore, several works have started to exploit large collections of process models with the aim of capturing a general understanding of how processes should be modeled or operated, with the aims of detecting deviating process instances~\cite{DBLP:journals/is/AaRL21,caspary2023does} and providing process modeling suggestions~\cite{sola2023activity}.

    Using LLMs, approaches to facilitating knowledge generation and maintenance, turning unstructured knowledge into executable specifications and queries, and tailoring these specifications and queries to a particular process context can be expected to become more effective and easier to implement.
    Hence, the application of LLMs to this end can be considered feasible and is expected to substantially impact BPM software in the near future. We also expect that the coming years of research and development will answer many nuanced open questions around LPM-augmented modeling and analysis capabilities, in particular about the interplay of auto-generated and hand-crafted symbolic models and meta-models, and the extent to which the importance of imperative models will decrease in favor of collections of declarative constraints that can be auto-tailored and assembled for modeling and analysis on demand, given the current context. Considering the prevalence of structured and unstructured process knowledge, e.g., in the form of the thousands of process models a single organization may own and the tens of thousands of process models that enterprise system vendors have at their hands, we expect that sufficient high-quality data exists to ``process fine-tune'' LLMs for generic and (if worth the cost) organization-specific BPM tasks and to provide high-value retrieval-augmented generation capabilities. \\
    \textbf{Challenges.} Given the short- to medium-term feasibility of Step~\ref{step1}, we assume that the key challenges that surround this step are of engineering nature. In particular, we consider the following challenges worth addressing:
    \begin{enumerate*}[i)]
    \item define rigid evaluation metrics for the generative AI-supported generation of process models and queries;
    \item specify and evaluate LLM-friendly data exchange formats that serve as ``middle-layer'' representation formats between symbolic and sub-symbolic sub-systems;
    \item assess the potential of fine-tuning LLMs for process model and query generation, as well as the potential of alternative or complementary approaches such as retrieval-augmented generation.
    \end{enumerate*}
 \item \textbf{Fusing unstructured and tabular data for actionable insights.}\label{step2}
 The previous step establishes the LPM as an augmentation of BPM, without changing BPM fundamentals. This step aims at utilizing the LPM to advance the frontier of business process analytics, particularly towards simulation and prediction.
 These capabilities have been the subject of comprehensive scientific studies and often utilize deep learning approaches.
 Neural network-based anomaly detection can allow organizations to infer actions that fix the identified anomalies, thus improving process performance~\cite{NOLLE2022101458}.
 Also, predictive monitoring approaches for business processes often utilize deep learning to predict future activities or outcomes or to classify cases~\cite{DBLP:journals/corr/abs-2101-09320}.
 Some simulation approaches utilize deep learning for generating more realistic business process simulation models, thus facilitating process improvements by enabling counterfactual (``what-if'') analyses~\cite{DBLP:conf/caise/CamargoDR22}.
 Finally, recent research even investigates the data-driven forecasting of entire process models~\cite{DBLP:journals/dke/SmedtYPWM23}.
 Despite these substantial research efforts, business process prediction and simulation tools are rarely applied at scale in industry and typically remain tools for basic exploration and not for high-impact analysis.
 Among the reasons for this are engineering challenges related to the training and re-training of highly specific (i.e.,\ organization- and process-specific) models at scale, the lack of holistic context in most event logs, the dynamism of business environments, and -- in the case of hybrid approaches such as process model-generation with deep learning -- the inability of traditional, symbolic process models to capture socio-organizational nuances. Utilizing foundation models can potentially both address the problem of lacking contextual knowledge by extracting this knowledge from unstructured or hard-to-search sources, and provide alternatives to simulation and prediction based on highly specific supervised training, by instead training foundation models on process execution traces that may be able to generalize simulation and prediction across process and organizational context (to a certain extent).
 However, due to the lack of research that systematically evaluates the potential of generative AI and foundation models in the aforementioned directions, feasibility remains an open question. \\
 \textbf{Challenges.} Our assessment is that Step~\ref{step2} requires the establishment of foundations that either already exist somewhat analogously for Step~\ref{step1} or that require a paradigm shift to facilitate fusing unstructured information and tabular data for insight generation. More specifically, we call for:
 \begin{enumerate*}[i)]
    \item establishing the conceptual foundations of conversational process mining, as well as of new, more applicable process simulation paradigms;
    \item devising and implementing approaches for eliciting serializable hypotheses about tabular data from unstructured information and informal knowledge;
    \item designing, implementing, and evaluating algorithms that allow for the evaluation of the aforementioned hypothesis in a scalable manner.
 \end{enumerate*}

 \item \textbf{Automating continuous improvement with the human in control.}\label{step3}
As the ultimate, long-term objective, LPMs can enable the automation of the BPM life-cycle -- i.e.,\ the continuous loop of process design, execution, analysis, and improvement -- with human involvement only for enabling full social control for key decision-making.
The question of whether this is, at all or to a certain extent, possible remains open.
Research on the (full) automation of the entire BPM lifecycle is scarce.
A notable line of work has proposed and evaluated the use of Developer Operations (DevOps) principles and practices in conjunction with reinforcement learning to this end~\cite{DBLP:journals/is/SatyalWPCM19}.
Here, the use of contextual multi-armed bandits is proposed to route process instances to the best possible process variant (configuration) given the particular case context.
With time, the contextual routing behavior is expected to converge, which can then trigger a final process change analogously to a change based on a classical A/B test.
The approach can be extended, to feature so-called \emph{shadow testing} that routes cases, in parallel to their actual execution, through hypothetical process variants, relying as much as possible on real-world properties and behavior and utilizing simulation only where necessary~\cite{DBLP:conf/otm/SatyalWPCM18}.
Shadow testing can then be used to narrow down the change candidates that are sufficiently promising for pilot tests.
The approach can be extended further to allow for human intervention, thus reducing the risk of unreasonable machine decisions given context that is available to a human expert but not to the machine~\cite{DBLP:conf/bpm/KurzSGKPW22,DBLP:conf/bpmds/KurzKPW23}.
From an industry perspective, the proposed approaches are very ambitious, as they require substantial flexibility and agility in the configuration of complex enterprise systems and primarily rely on process performance data when making decisions about process changes.
Foundation models can potentially allow for fusing insights based on execution data with structured and unstructured knowledge, while also maintaining a reinforcement learning-like feedback loop that continuously re-evaluates generated insights and actions.
Beyond that, LPMs can also help utilize data from poorly structured (sub-)processes; for example, many hiring processes mostly take place on a social level and leave a trace of textual information that is difficult to analyze with traditional process mining approaches; this leaves a gap that LLM-augmented business process analytic can potentially fill.
Hence, LPMs as bridges between data-driven, knowledge-intense, and social decision-making may enable a leap forward to more machine autonomy on the level of the BPM life-cycle.
However, considering the scarcity of related research, general feasibility remains an open question, in particular when considering complexity in the context of traditional enterprise systems (time between action and effect as well as the size of the action-space), as well as reliability and compliance requirements for high impact process changes. \\
\textbf{Challenges.} Because Step~\ref{step3} requires the fusion of management and execution, both on a technical and organizational level, we consider it substantially more challenging than Step~\ref{step2}, which is largely independent from execution systems.
In addition, we assume that dynamic process adjustment to specific organizational context will, for the foreseeable future, always require a trade-off with standardization, e.g., to reign in the costs and risks associated with organizational complexity.
Accordingly, we pose challenges in the form of the following questions:
 \begin{enumerate*}[i)]
    \item how can enterprise software, beyond process execution engines, be designed for maximal flexibility and modularity on the \emph{knowledge level}, such that desired changes to a process or process variant can be deployed with minimal human involvement?
    \item how do the fundamental requirements to analysis-oriented BPM software (e.g., to process mining software) change when the software becomes a mission-critical component of execution systems?
    \item how can guardrails for (somewhat) autonomous process execution systems be defined and their compliance ensured, so that \emph{standardization versus tailored optimization} trade-offs can be shifted from the former towards the latter?
\end{enumerate*}
\end{enumerate}
Feasibility challenges that are orthogonal to the three steps above relate to data management, reliability \& compliance, and interaction of human and machine decision-making:
\begin{itemize}
    \item The LPM consumes data from a broad range of sources and processes it in various ways so that it can be used efficiently by humans and machines in the BPM lifecycle. Integrating with external data sources and managing the ingested and generated data is a key challenge, particularly because LLM-generated data and knowledge may be of questionable quality and require substantial curation, either by more reliable machines or humans. Hence, one risk of the LPM proposal and similar deployments of generative AI is that the well-known problems pertaining to data management and BPM (arguably most pronounced in the context of extract-transform-load pipelines of process mining~\cite{DBLP:conf/bpmds/KampikW22}) will be further exacerbated, thus requiring innovation in the sub-field of process querying methods~\cite{DBLP:books/sp/22/P2022}. The aforementioned data management challenges are generally well-known in applied AI research and have led to the emergence of \emph{data-centric AI} -- an engineering paradigm that focuses on data management and data pipelines as key foundations of ML-based applications~\cite{DBLP:journals/corr/abs-2211-05764}.
    \item LLMs are frequently criticized for the lack of reliability of the output they produce and have been described by experts as \emph{stochastic parrots}~\cite{10.1145/3442188.3445922} and \emph{bullshit generators}\footnote{See: \url{https://www.aisnakeoil.com/p/chatgpt-is-a-bullshit-generator-but}, \emph{accessed at 28-03-2023}. Technically, \emph{bullshit} is a statement that is uttered by an agent with indifference to the statement's truth~\cite{Frankfurt2005}.}. Hence, it is crucial that insights and actions inferred by LLMs and other deep learning models are automatically assessed regarding their reliability and their business and societal implications, such as fairness~\cite{DBLP:journals/corr/abs-1908-11451}. Beyond that, a key issue is that the ingestion of further content will increase the ethics and compliance risk of personal information leakage, a problem that has recently sparked substantial research interest in the context of process mining~\cite{DBLP:journals/dke/FahrenkrogPetersenAW23,DBLP:journals/is/FahrenkrogPetersenKAW23}.
    Potential ethics and privacy issues go hand-in-hand with requirements to ensure legal compliance, which traditionally is a challenge that BPM aims to address~\cite{compliance} and not to exacerbate.
    \item Even if the inferences drawn by the LPM (or: an underlying LLM) are technically verifiable, they may still pose challenges to human decision-making. For example, a query or process configuration specification may be technically correct and human-interpretable but require substantial cognitive effort to make sense of; if the LPM then recommends the execution of the query or specification to a human user, the user may trigger the execution without carefully checking, not detecting flaws that could have been identified only with human knowledge that is not maintained on a purely technical level. The more severe the consequence of an action recommended or influenced by an LPM is, the more important it is that human experts carefully deliberate the action's implications before executing or triggering it. Here, concepts from behavioral psychology such as \emph{choice architecture}~\cite{thaler2013choice} that study how human decision-making is influenced by contextual information can be utilized, which have already been adopted by the information systems realm~\cite{DBLP:journals/bise/WeinmannSB16}.
    \item The limits of economic feasibility of LLM (and more broadly: foundation model) training, operation, and maintenance are a moving frontier. It is well-known that foundation model training, and hence also full-blown updates of foundation models, are very costly (in the millions of USD). Even drawing inferences from pre-trained models can incur substantial costs, surpassing the costs of operating traditional symbolic or statistical inference systems. Hence, for each application of foundation models (and, as a consequence, of LPMs) it is crucial to assess whether the costs exceed the benefits and whether alternative technologies may achieve better scores in a cost/benefit calculation. For example, in some use cases, utilizing the smaller pre-trained natural language processing models of popular Python libraries for semantic similarity matching may make more sense than relying on a more costly fine-tuned LLM, which always entails a lock-in to the specific LLM architecture/model. A compromise between the two options may be a powerful general-purpose LLM that heavily relies on retrieval augmented-generation, thus utilizing the potential of existing knowledge-based systems.
    Beyond that, deploying smaller models that are trained based on feedback from very large models has emerged as a promising research direction~\cite{mukherjee2023orca}, which could potentially facilitate the more cost-efficient use of foundation models.
\end{itemize}
In all three cases, the challenges are reasonably feasible to address for Step~\ref{step1}: here, the generation of symbolic knowledge (i.e., models and queries) can be managed using well-established data- and knowledge-base technologies, is verifiable, and can be wrapped into user-friendly abstractions in relatively straightforward and well-understood procedures.
In contrast, both Step~\ref{step2} and Step~\ref{step3} pose substantial challenges regarding the management of ML models, such as a potential process execution traces-based foundation model, as well as regarding the variability of results such as predictions and action recommendations.

\section{Discussion}
\label{sec:discussion}
This section relates the proposal to other visions and overviews of generative AI and BPM and briefly discusses BPM and generative AI for modalities other than text.
\subsection{Related Concepts}
Considering the current hype around LLMs and generative AI, conceptual proposals and implementations for the domain-specific use of LLMs emerge at a fast pace.
One prominent example is the development of \emph{BloombergGPT}, a special-purpose LLM trained specifically for the finance domain~\cite{wu2023bloomberggpt}.
Unsurprisingly, the first comprehensive proposals for fusing BPM and LLMs have emerged as well.
Notably, Vidgof et al. lay out a vision and research agenda for LLMs and BPM~\cite{DBLP:journals/corr/abs-2304-04309}; their work is primarily aligned with the BPM life-cycle, i.e., we claim that it provides a management view on LLMs for BPM, whereas our perspective is feasibility-oriented.
Beheshti et al. propose \emph{ProcessGPT}~\cite{beheshti2023processgpt}, a transformer-based approach for recommending next actions in knowledge-intensive processes during execution\footnote{Note that we consider one of the proposed use cases -- automated exam grading and plagiarism detection -- to be highly questionable from ethics and feasibility perspectives.}. Analogously to BloombergGPT, ProcessGPT is envisioned as a special-purpose GPT, trained from scratch with domain-specific data.
Hence, the difference to our LPM proposal is two-fold: \emph{i)} our scope is broader, encompassing the entire BPM life-cycle and \emph{ii)} we do not primarily propose training a GPT from scratch, under the assumption that the costs out-weigh the benefits and that fine-tuning and prompt-based contextualization are better means for reaching the same objective in the context of large language models. While training foundation models on process data for prediction and counterfactual simulation purposes is part of the LPM research agenda, the general feasibility of LPMs as a broader approach is not dependent on the feasibility of this particular potential capability.
Focusing on process data analysis, Berti and Qafari~\cite{berti2023leveraging} propose approaches for utilizing off-the-shelf LLMs for process mining, in particular for directly answering user queries and for generating symbolic queries on process data. The proposals are supported by preliminary experiments, providing evidence of feasibility.
Given the (smaller) scope of the paper by Berti and Qafari, we consider the approach proposed in their work as a subset of the capabilities of what LPMs can offer, providing first and partial evidence for LPM feasibility.
Similarly, we see the works by Klievtsova et al.~\cite{klievtsova2023conversational} and Grohs et al.~\cite{DBLP:journals/corr/abs-2307-09923} as conceptual and experimental starting points for LPMs for process modeling.
Here, we can again highlight that what is still missing are experimental works that provide solid evidence for the effectiveness of LLMs in a process execution context.

\subsection{Generative AI for BPM beyond LLMs}
The LPM proposal and its LLM analogy place text-based generative AI into the center of attention.
Beyond this, foundation models specifically trained on process execution traces may be utilized by the LPM for prediction and simulation.
Obviously, other modalities such as image, video, and sound are relevant as well.
For example, process models are often created as part of notorious slide decks, making it harder to govern the models and utilize them for data analysis. To make it easier to move from images to formal process model representations, recent research introduces a deep learning-based approach
to turn images of process flow into standard-compliant (XML-based) BPMN~\cite{DBLP:journals/tse/SchaferALS23}.
In this context, one could imagine that generative AI can be applied, if not directly as image processors, then as post-processors of the XML output.
Also, generative AI models could potentially be applied to automatically generate insights, such as models and database queries, from large amounts of collected audio data, such as from expert interviews or customer conversations.
However, here it is again not clear whether the additional modality (sound) is best to be processed directly by a foundation model; pre-processing with an off-the-shelf speech-to-text processor may be more feasible and easier to deploy.

In conclusion, it must be noted that our LPM vision is primarily presented with regard to the current state of BPM and business processes considering the ongoing discussion about how generative AI can be generally applied in business processes. However, it remains uncertain how, or even if, generative AI will bring about fundamental changes in BPM practices or business processes that require fundamental changes in BPM approaches, such as the BPM lifecycle, from a management perspective.

\section{Conclusion}
\label{sec:conclusion}
In this paper, we have introduced the notion of a Large Process Model (LPM) that allows for the automated inference of insights and actions with respect to a specific process in a given organizational context based on a large and heterogeneous collection of data and knowledge about many processes across many organizational contexts, with the goal of facilitating BPM now and in the future in light of advancements in generative AI.
While our LPM utilizes a (foundation model-based) LLM and potentially process execution data-specific foundation models, we see the LPM as a fusion of generative AI and traditional symbolic and statistical approaches to automating reasoning and decision-making in BPM.
We assess the application of process fine-tuned general purpose LLMs as contextualizers, generators, and augmenters of symbolic models and queries as feasible and as substantial facilitators of BPM.
Here, we expect a substantial industry impact over the next years.
Beyond that, we view the usage of special-purpose foundation models for BPM, in particular based on process execution traces, as a promising research frontier but as too nascent to warrant predictions of large-scale industry deployments.
Further, the application of generative AI for automating larger parts of the BPM life-cycle is potentially interesting as well, but poses substantial feasibility challenges and business/societal risks that require extensive research and validation before a potential deployment is viable.

\begin{acknowledgements}
The authors would like to thank the numerous colleagues in academia and industry who work at the application of LLMs and generative AI and whose work has influenced the perspective provided in this paper.
\end{acknowledgements}

%
\section*{Conflict of interest}
All authors with a non-academic affiliation work for a company that builds business process management software.

\bibliographystyle{spmpsci}      

\end{document}